\documentclass[preprint,aps,floats,showpacs]{revtex4}
\usepackage{graphicx}
\usepackage{epsfig}
\usepackage{pstricks}
\usepackage{pst-coil}
\usepackage{amsmath}
\def\be{\begin{equation}}
\def\ee{\end{equation}}
\def\bea{\begin{eqnarray}}
\def\eea{\end{eqnarray}}

{\newcommand{\lsim}{\mbox{\raisebox{-.6ex}{~$\stackrel{<}{\sim}$~}}}
{\newcommand{\gsim}{\mbox{\raisebox{-.6ex}{~$\stackrel{>}{\sim}$~}}}

\def\ev{\,{\rm eV}}
\def\gev{\,{\rm GeV}}

\begin{document}
\title{\bf $B-L$ Cosmic strings and Baryogenesis}
\author{Pijushpani Bhattacharjee$^b$}
\email{pijush@iiap.res.in} 
\author{Narendra Sahu$^a$}
\email{narendra@phy.iitb.ac.in}
\author{Urjit A.~ Yajnik$^a$}
\email{yajnik@phy.iitb.ac.in}
\affiliation{$^a$Department of Physics, Indian Institute of
Technology, Bombay, Mumbai 400076, India} 
\affiliation{$^b$Indian Institute of Astrophysics, 
Bangalore-560 034, India}                            
%\date{}
\pagestyle{empty}
\begin{abstract}
\noindent 
Cosmic strings arising from breaking of the $U(1)_{B-L}$ 
gauge symmetry that occurs in a wide variety of unified 
models can carry zero modes of heavy Majorana neutrinos.
Decaying and/or repeatedly self-interacting closed loops 
of these ``$B-L$'' cosmic strings can be a non-thermal 
source of heavy right-handed Majorana neutrinos whose decay 
can contribute to the observed baryon asymmetry of the 
Universe (BAU) via the leptogenesis route. The $B-L$ cosmic 
strings are expected in GUT models such as $SO(10)$, where 
they can be formed at an intermediate stage of symmetry 
breaking well below the GUT scale $\sim 10^{16}$ GeV; such 
light strings are not excluded by the CMB anisotropy data 
and may well exist. We estimate the contribution of $B-L$ 
cosmic string loops to the baryon-to-photon ratio of the 
Universe in the light of current knowledge on neutrino masses 
and mixings implied by atmospheric and solar neutrino 
measurements. We find that $B-L$ cosmic string loops can contribute 
significantly to the BAU for $U(1)_{B-L}$ symmetry
breaking scale $\eta_{B-L}\gsim 1.7\times 10^{11}\gev$. 
At the same time, in order for the contribution of 
decaying $B-L$ cosmic string loops not to exceed the 
observed baryon-to-photon ratio inferred from the recent WMAP 
results, the lightest heavy right-handed Majorana neutrino 
mass $M_1$ must satisfy the constraint $M_1 \leq 2.4 \times 
10^{12}\left(\eta_{B-L}/10^{13}\gev\right)^{1/2}\gev$.  
This may have interesting implications for the 
associated Yukawa couplings in the heavy neutrino sector and 
consequently for the light neutrino masses generated through 
see-saw mechanism.  

\end{abstract}
\maketitle
%%%%%%%%%%%%%%%%%%%%%%%%%%%%%%%%%
%\newpage
%\pagenumbering{arabic}
\section{Introduction}
%%%%%%%%%%%%%%%%%%%%%%%%%%%%%%%%%%
A very attractive scenario of origin of the baryon ($B$) 
asymmetry of the Universe (BAU) is that it arose from a 
lepton ($L$) asymmetry~\cite{fukugita.86,luty.92,plumacher.96}. 
The conversion of the $L$-asymmetry to the $B$-asymmetry occurs 
via the high temperature behavior of the $B+L$ anomaly of the 
Standard Model~\cite{krs.86}. This is an appealing route for 
several reasons. First, the extremely small neutrino masses, 
suggested by the atmospheric neutrino-~\cite{atmos_data}, solar 
neutrino-~\cite{solar_data} and KamLAND experiment~\cite
{kamland_data} data, point to the possibility of Majorana 
masses for the neutrinos; such small neutrino mass can be 
generated, for example, through the see-saw mechanism~\cite
{mohapatra-pal-book} that involves heavy right-handed neutrinos 
whose interactions involve $L$ violation in a natural way. Second, 
most particle physics models incorporating the above possibility 
demand new Yukawa couplings and also possibly scalar
self-couplings; these are the kind of couplings which, unlike 
gauge couplings, can naturally accommodate adequate $CP$ 
violation, one of the necessary ingredients~\cite{sakharov.67} for 
generating the BAU. 

Most proposals along these lines rely on out-of-equilibrium
decay of the thermally generated right-handed heavy Majorana 
neutrinos in the early Universe to generate the $L$-asymmetry. 
The simplest possibility to implement this scenario is 
to extend the Standard Model (SM) by the inclusion of a 
right handed neutrino, $\nu_R$. A more appealing alternative is to 
consider this within the context of unified models with an embedded   
$U(1)_{B-L}$ gauge symmetry. For example, it can be the Left-Right 
symmetric model~\cite{rabi&jogesh.75, rabi.81} where $B-L$ is 
naturally required to be a gauge charge, or it can be a Grand 
Unified Theory (GUT) based on $SO(10)$ gauge group. Because 
$B-L$ is a gauge charge in such models, no primordial $B-L$ 
can exist as long as the $U(1)_{B-L}$ gauge symmetry remains 
unbroken. However, spontaneous breaking of the $U(1)_{B-L}$ 
gauge symmetry gives heavy Majorana mass to the right-handed 
neutrinos, and a net $B-L$ can be dynamically generated through 
out-of-equilibrium decay of these heavy right-handed Majorana 
neutrinos. Rapid violation of $B+L$ by the high temperature 
sphaleron transitions erases any $B+L$ generated earlier. 
These sphaleron transitions, however, conserve $B-L$. Thus, 
in this scenario the final BAU is related to the $B-L$ produced 
after the $U(1)_{B-L}$ symmetry breaking phase transition. 

One of the interesting features of any $U(1)$ gauge symmetry 
breaking phase transition in the early universe is the possible 
formation of cosmic strings~\cite{kibble.76,vilen&shell}. It 
has been noted earlier by several authors~\cite{pijush.82, 
brandenberger&co.91,riotto&lewis.94,jeannerot.96, pijush.98} 
that decaying, collapsing, or repeatedly self-intersecting  
closed loops of such cosmic strings, can be a {\it non-thermal} 
source of massive particles that ``constitute'' the string, 
and that the decay of these massive particles can give rise 
to the observed BAU or at least can give significant 
contribution to it. Cosmic strings formed at a phase 
transition can also influence the nature of a subsequent phase 
transition that may have important implications for the 
generation of BAU~\cite {sbdanduay1, brandenetal}.

In the present context, the ``$B-L$'' cosmic strings associated 
with the $U(1)_{B-L}$ symmetry breaking phase transition 
mentioned above are of particular interest~\cite{jeannerot.96} 
because they can carry zero modes~\cite{jackiw.81,weinberg.81} 
of the heavy right-handed neutrinos $\nu_R$. This is possible 
because the higgs field involved in the cosmic string solution 
arising from the spontaneous breaking of the $U(1)_{B-L}$ is 
the same higgs that gives heavy Majorana mass to the $\nu_R$ 
through Yukawa coupling. It has, therefore, been suggested~\cite{
jeannerot.96} that decaying closed loops of such cosmic strings 
can be an additional, non-thermal, source of the $\nu_R$'s, 
whose subsequent decay can contribute to the BAU through the 
leptogenesis route. 

In this paper we revisit this scenario of generating the BAU 
through the decay of $\nu_R$'s released from $B-L$ cosmic 
string loops in the light of recent ideas about neutrino 
masses and mixings implied by the solar and atmospheric 
neutrino data. We find that, $B-L$ cosmic string loops can 
contribute significantly to the BAU for $U(1)_{B-L}$ symmetry
breaking scale $\eta_{B-L}\gsim 1.7\times 10^{11}\gev$.
At the same time we show that, in order for the contribution 
of decaying $B-L$ cosmic string loops not to exceed the 
observed baryon-to-photon ratio inferred from the recent WMAP 
results~\cite{wmap}, the lightest heavy right-handed Majorana 
neutrino mass $M_1$ must satisfy the constraint $M_1\leq 2.4 
\times10^{12}\left(\eta_{B-L}/10^{13}\gev\right)^{1/2}\gev$. 

The rest of this paper is organized as follows. In section 
II we briefly discuss some examples of symmetry breaking schemes 
in unified models with an embedded gauged $U(1)_{B-L}$ which 
potentially allow cosmic string solutions, and discuss an 
explicit example of a cosmic string solution in the context 
of a simple extension of the SM, namely, the gauge group $SM 
\otimes U(1)_{B-L}$ spontaneously broken to SM. The nature of 
the neutrino zero modes in presence of such a cosmic string is 
then discussed. In section III we briefly review the evolution of 
cosmic strings with particular attention to formation of 
closed loops and their subsequent evolution, and the production 
of massive particles from decaying and/or repeatedly 
self-intersecting cosmic string loops. We also discuss the 
observational constraints on the relevant cosmic string 
parameters. We then estimate, in section IV, the contribution 
of the $B-L$ cosmic string loops to the BAU, and discuss the 
constraint on the lightest heavy right-handed Majorana neutrino
mass $M_1$. Section V concludes the paper with a brief summary 
of our main results. Throughout this paper we use natural units 
with $\hbar=c=k_B=1$. 
%%%%%%%%%%%%%%%%%%%%%%%%%%%%%%%%%%
\section{$U(1)_{B-L}$ cosmic strings and neutrino zero modes}
%%%%%%%%%%%%%%%%%%%%%%%%%%%%%%%%%
There are several realistic particle physics models 
where a gauged $B-L$ symmetry exists and breaks at a certain 
scale. Since $SO(10)$ minimally incorporates $B-L$ 
gauge symmetry we consider the models embedded in 
$SO(10)$. The following breaking schemes can 
potentially accommodate cosmic strings. One of the 
breaking schemes, motivated by supersymmetric 
$SO(10)$~\cite{jeannerot&davis.95,jeannerot_prd.96}, involves the 
intermediate 
left-right symmetric model: 
\bea
SO(10) &\underrightarrow{54+45}& SU(3)_{C}\otimes SU(2)_{L}\otimes 
SU(2)_{R}\otimes U(1)_{B-L}\nonumber\\
&\underrightarrow{126+\overline{126}}& SU(3)_{C}\otimes 
SU(2)_{L}\otimes U(1)_{Y} \otimes Z_{2}\nonumber\\
&\underrightarrow{10+10^{'}}& SU(3)_{C}\otimes U(1)_{Q}\otimes 
Z_{2}\,.
\eea 
During the first phase of symmetry breaking, presumably at a  
GUT scale of $\sim10^{16}\gev$, monopoles 
are formed. However, during the second and third phases 
of symmetry breaking cosmic strings are formed since 
$\pi_{1}(\frac{3221}{321})=\pi_{1}(\frac{321}{31})=Z_{2}$,  
where the numbers inside the parentheses symbolize the 
group structures. The monopole problem in this model 
can be solved by using a hybrid inflation ending at the 
left-right symmetric phase of the Universe~\cite{jeannerot_prd.96} 
thus inflating away the monopoles. The formation of cosmic 
strings in the later phases is of great interest since 
these ``light'' (i.e., lighter than GUT scale) cosmic strings 
do not conflict with any cosmological observations. The $Z_{2}$ 
strings of the low energy theory was investigated in an 
earlier work~\cite{yajnik.99}.

Another scheme is to break supersymmetric $SO(10)$ 
directly to $SU(3)_{C}\otimes SU(2)_{L}\otimes U(1)_{R}\otimes 
U(1)_{B-L}$ with the inclusion of extra $54+45$. The 
first $45$ acquires a vacuum expectation value along 
the direction of $B-L$. However, the latter $45$ acquires 
a vacuum expectation value along the direction of $T_{3R}$:    
\bea
SO(10) &\underrightarrow{54+45+54^{'}+45^{'}}& SU(3)_{C}\otimes 
SU(2)_{L}\otimes U(1)_{R}\otimes U(1)_{B-L}\nonumber\\
&\underrightarrow{126+\overline{126}}& SU(3)_{C}\otimes 
SU(2)_{L}\otimes U(1)_{Y} \otimes Z_{2}\nonumber\\
&\underrightarrow{10+10^{'}}& SU(3)_{C}\otimes U(1)_{Q}\otimes Z_{2}
\eea

For the present purpose it is sufficient to consider a 
model based on the gauge group $SM \otimes U(1)_{B-L}$ which is 
spontaneously broken to SM. 
Existence of cosmic strings and the related zero-modes
in this model can be established as follows. Let the gauge field 
corresponding to the $U(1)_{B-L}$ symmetry be denoted by $C_{\mu}$, 
and the symmetry be broken by a SM singlet
$\chi$. Let $\langle\chi\rangle$ be $\eta_{B-L}$ below the critical 
temperature $T_{B-L}$. In a suitable gauge a long cosmic string
oriented along the $z$-axis can be represented (in cylindrical polar 
coordinates) by the ansatz~\cite{vilen&shell}
\begin{eqnarray}
\chi&=&\eta_{B-L} f(r)e^{in{\theta}} \label{string_higgs}\,,\\
C_{\mu}&=&\frac{ng(r)}{\alpha r}\delta^{\theta}_{\mu}\,,
\label{string_gauge}
\end{eqnarray}
where $n$ is an integer giving the winding number of the phase of the 
complex higgs field $\chi$, and $\alpha$ is the gauge coupling constant 
for the group $U(1)_{B-L}$. 
In order for the solution to be regular at the origin we 
set $f(0)=g(0)=0$. Also requiring the finiteness of energy of 
the solution, we set $f(r)=g(r)=1$ as $r\rightarrow\infty$. 
It turns out that both $f(r)$ and $g(r)$ take their asymptotic
values everywhere outside a small region of the order of 
$\eta_{B-L}^{-1}$ around the string. Thus away from
the string $\langle\chi\rangle=\eta_{B-L} $ up to a phase, and $C_{\mu}$ 
is a pure gauge. The mass scale of the string is fixed by 
the energy scale of the symmetry breaking phase transition 
$\eta_{B-L}$ at which the strings are formed. Then the mass 
per unit length of a cosmic string, $\mu$, is of order 
$\eta^{2}_{B-L}\sim T^{2}_{B-L}$. 

The lagrangian for the right-handed neutrino is 
\be
\mathcal{L}_{\nu_R}= i\overline{\nu_{R}}\sigma^{\mu}
D_{\mu}\nu_{R} -{1\over 2}
[ih\overline{\nu_{R}}\chi\nu_{R}^{c}+H.C]\,,
\label{lagrangian_nu_R}
\ee
where $h$ is the Yukawa coupling constant, $\sigma^{\mu}=(-I, 
\sigma^{i})$, and $\nu_{R}^{c}=i \sigma^{2}\nu_{R}^{*}$ defines 
the Dirac charge conjugation operation. The resulting equations 
of motion have been shown~\cite{jackiw.81} to possess $|n|$ 
normalisable zero-modes in winding number sector $n$. The field 
equations in the  $U(1)$ example
are~\cite{sdavis.00}
\be
\begin{pmatrix} -e^{i\theta}[\partial_{r}+{i\over r}
\partial_{\theta}+{ng(r)\over 2r}] & \partial_{z}+\partial_{t}\\
\partial_{z}-\partial_{t} & e^{-i\theta}[\partial_{r}-{i\over r}
\partial_{\theta}-{ng(r)\over 2r}]\end{pmatrix}\nu_{R}-M_{R}
e^{in\theta}\nu_{R}^{*}=0\,,
\label{field_equations_nu_R}
\ee
where the expressions (\ref{string_higgs}) and (\ref{string_gauge}) 
have been substituted for $\chi$ and $C_{\mu}$, and 
$M_{R}=h\eta_{B-L}$. In the winding number sector $n$ the 
normalisable zero-modes obey $\sigma^{3}\psi=\psi$ and are 
of the form 
\be
\nu_{R}(r, \theta)=\begin{pmatrix}1\\
0\end{pmatrix}\left(U(r)e^{il\theta}+V^{*}(r)e^{i(n-1-l)
\theta}\right)\,,
\label{zeromode}
\ee                                                                   
where $U(r)$ and $V(r)$ are well behaved functions at the 
origin and have the asymptotic behavior $\sim \exp(-M_{R}r)/
\sqrt{r}$. When nontrivial $z$ and $t$ dependences are included,
these modes have solutions that depend on $z+t$ and are Right movers. 
For $n<0$, normalisable solutions obey $\sigma^{3}\psi=-\psi$,
and form the zero-energy set of a Left moving spectrum. On a straight 
string these modes are massless. However on wiggly strings 
they are expected to acquire effective masses proportional 
to  the inverse radius of the string curvature.  

%%%%%%%%%%%%%%%%%%%%%%%%%%%%%%%%%%%%%%%%%
\section{Evolution of cosmic strings: Formation and evolution 
of closed loops and production of massive particles}
%%%%%%%%%%%%%%%%%%%%%%%%%%%%%%%%%%%%%%%%%%
\subsection{Scaling solution and closed loop formation} 
The evolution of cosmic strings in the expanding Universe has 
been studied extensively, both analytically as well as 
numerically; for a text-book review, see the 
monograph~\cite{vilen&shell}. Here we briefly summarize 
only those aspects of cosmic string evolution that are 
relevant for the subject of the present paper, namely the 
formation and subsequent evolution of closed loops 
of strings and production of massive particles from them. 
This closely follows the discussion in section 6.4 of 
Ref.~\cite{pijushreport}. 

Immediately after their formation at a phase transition, the 
strings would in general be in a random tangled configuration. 
One can characterize the string configuration in terms of a 
coarse-grained length scale  $\xi_s$ such that the overall 
string energy density $\rho_s$ is given by $\rho_s=\mu/\xi_s^2$. 
Initially, the strings move under a strong damping force due 
to friction with the ambient thermal plasma. In the friction 
dominated epoch a curved string segment of radius of curvature 
$r$ acquires a terminal velocity $\propto 1/r$. As a result the 
strings tend to straighten out so that the total length of 
the strings decreases. Thus the overall energy density in the form 
of strings decreases as the Universe expands. This in turn means 
that the length scale $\xi_{s}$ increases. Eventually, $\xi_s$ 
reaches the causal horizon scale $\sim t$. After the damping 
regime ends (when the background plasma density falls to a 
sufficiently low level as the Universe expands), the strings 
start to move relativistically. However, causality prevents the 
length scale $\xi_{s}$ from exceeding the horizon size $\sim 
t$. Analytical studies supported by extensive numerical simulations 
show that the subsequent evolution of the system is such that 
the string configuration reaches a ``scaling regime'' in which 
the ratio $\frac{\xi_{s}}{t}\equiv x$ remains a constant. 
Numerical simulations generally find the number $x$ to lie 
approximately in the range $\sim$ 0.4--0.7. This is called 
the scaling regime because then the energy density in the form 
of strings scales as, and remains a constant fraction of, the 
energy density of radiation in the radiation dominated epoch or 
the energy density of matter in the matter dominated epoch both 
of which scale as $t^{-2}$. 

The fundamental physical process that maintains the string 
network in the scaling configuration is the formation of 
{\it closed loops} which are pinched off 
from the network whenever a string segment curves over into a 
loop, intersecting itself. In the ``standard'' 
picture~\cite{vilen&shell}, the closed loops so formed 
have average length at birth    
\be 
L_b=K\Gamma G\mu t\,,
\label{birth_length}
\ee 
and they are formed at a 
rate (per unit volume per unit time) which, in the radiation 
dominated epoch, is given by 
\be
\frac{dn_{b}}{dt}=\frac{1}{x^{2}}\left(\Gamma G \mu\right)
^{-1} K^{-1}t^{-4}\,,
\label{birth_rate}
\ee
where $\Gamma\sim 100$ is a geometrical factor that determines the 
average loop length, and $K$ is a numerical factor of order unity. 

The whole string network consisting of closed loops as well as 
long strands of strings stretched across the horizon gives rise 
to density fluctuations in the early Universe which could 
potentially contribute to the process of formation of structures 
in the Universe. More importantly, they would produce specific 
anisotropy signatures in the cosmic microwave background (CMB). 
Using a large-scale cosmic string network simulation and comparing 
the resulting prediction of CMB anisotropies with observations, 
a recent analysis~\cite{landriau-shellard} puts an upper limit 
on the fundamental cosmic string parameter $\mu$, giving 
$G\mu\lsim 0.7\times 10^{-6}$. This translates to an upper limit, 
$\eta\lsim 1.0\times10^{16}\gev$, on the symmetry-breaking 
energy scale of the cosmic string-forming phase transition. This 
probably rules out cosmic string formation at a typical GUT 
scale $\sim10^{16}\gev$. However, lighter cosmic strings arising 
from symmetry breaking at lower scales, such as the $B-L$ cosmic 
strings in the case of the $SO(10)$ model discussed in the 
previous section, are not ruled out. 

It should be noted here that, in the standard scenario of cosmic 
string evolution described above, the loops are formed on a length scale 
that is a constant fraction of the horizon length, as given by equation 
(\ref{birth_length}). Thus, the average size of the newly formed loops 
increases with time. At the relevant times of interest, these loops, 
although small in comparison to the horizon scale, would still be of 
macroscopic size in the sense that they are much larger than the 
microscopic string width scale $w\sim\eta^{-1}\sim\mu^{-1/2}$. 

In contrast, results of certain Abelian Higgs (AH) model 
simulations of cosmic string evolution~\cite{vah} seem to 
indicate that scaling configuration of the string network is maintained 
primarily by loops formed at the smallest fixed length scale in the 
problem, namely, on the scale of the width $w\sim\eta^{-1}\sim\mu^{-1/2}$ 
of the string. These microscopic ``loops''  
quickly decay into massive particles (quanta of gauge bosons, higgs 
bosons, heavy fermions etc.) that ``constitute'' the string. In other 
words, in this scenario, there is essentially no macroscopic loop 
formation at all; instead, the scaling of the string network is maintained 
essentially by massive particle radiation. In order for the scaling 
configuration of the string network to be maintained by this process, the 
microscopic loops must be formed at a rate  
\be
\left(\frac{dn_{b}}{dt}\right)_{\rm AH}=\frac{1}{x^{2}}
\mu^{1/2}t^{-3}\,.
\label{birth_rate_AH}
\ee

The above scenario of cosmic string evolution in which massive particle 
radiation rather than gravitational radiation plays the dominant role  
is, however, currently a subject of debate~\cite{moore-shell}. One of the 
major problems hindering a resolution of the issues involved is the 
insufficient dynamic range possible in the currently 
available AH model simulations and the consequent need for extrapolation 
of the simulation results to the relevant cosmological scales, 
which is not straightforward. In this paper, we shall primarily restrict 
ourselves to consideration of the ``standard'' macroscopic loop formation 
scenario described by equations (\ref{birth_length}) and 
(\ref{birth_rate}) above, although we shall have occasions to refer to the 
massive particle radiation scenario below (see, in particular, section 
III.B.2).     

\subsection{Fate of the closed loops and massive particle production} 
The behavior of the closed loops after their formation may be 
broadly categorized into following two classes:  

\subsubsection{Slow death} 
Any closed loop of length $L$ in its center of momentum frame 
has an oscillation period $L/2$~\cite{kibble&turok-82}. However, 
a loop may be either in a self-intersecting or non-selfintersecting 
configuration. In general, a closed loop configuration can be 
represented as a superposition of waves consisting of various 
harmonics of $\sin$'s and $\cos$'s. Some explicit low harmonic 
number analytical solutions of the equations of motion of 
closed loops representing non-selfintersecting loops are known 
in literature~\cite{kibble&turok-82, turok-84, chendicarlo&hotes-88,
delaneyetal-90}, and it is possible that there exists a large 
class of such non-selfintersecting solutions. Indeed, numerical 
simulations, while limited by spatial resolution, do seem to 
indicate that a large fraction of closed loops are born in 
non-selfintersecting configurations. 

A non-selfintersecting loop oscillates freely. As it oscillates, 
it loses energy by emitting gravitational radiation, and 
thereby shrinks. When the radius of the loop becomes of the 
order of its width $w\sim\eta^{-1}\sim\mu^{-1/2}$, the 
loop decays into massive particles. Among these particles will be the 
massive gauge bosons, higgs bosons, and in the case of the 
$B-L$ strings, massive right-handed neutrinos ($\nu_R$) which 
were trapped in the string as fermion zero modes. We shall 
hereafter collectively refer to all these particles as $X$ 
particles. We are, of course, interested here only in the $\nu_R$'s. In 
addition to those directly released from the loop's final decay, 
there will also be some $\nu_R$'s coming from the decays of the gauge and 
higgs bosons released in the final loop decay.  
It is difficult to calculate exactly the total number of $\nu_R$'s so 
obtained from each loop, but we may expect that it would be a number of 
order unity. For the purpose of this paper we shall assume that each final 
demise of a loop yields a number $N_N\sim O(1)$ of heavy right handed 
Majorana neutrinos; we shall keep this number $N_N$ as a free parameter in 
the problem. 

The rate of release of $\nu_R$'s at any time $t$ by the above 
process can be calculated as follows. The lifetime of a loop of 
length $L$ due to energy loss through gravitational wave radiation is 
\be
\tau_{\rm GW}\sim \left(\Gamma G \mu\right)^{-1}L\,.
\label{gw_lifetime}
\ee
Equations (\ref{birth_length}) and (\ref{gw_lifetime}) thus show that 
loops born at time $t$ have a lifetime $\sim Kt\gsim H^{-1}(t)$, 
where $H^{-1}(t)\sim t$ is the Hubble expansion time scale. It is 
thus a slow process. From the above, we see that the loops that are 
disappearing at any time $t$ are the ones that were 
formed at the time $(K+1)^{-1}t$. Taking into account the 
dilution of the number density of loops due to expansion of 
the Universe between the times of their birth and final demise, 
equation (\ref{birth_rate}) gives the number of loops disappearing 
due to this ``slow death'' (SD) process per unit time per unit 
volume at any time $t$ (in the radiation dominated epoch) as  
\be
\frac{dn_{\rm SD}}{dt}=f_{\rm SD}\frac{1}{x^{2}}\left(\Gamma G \mu\right)
^{-1} \frac{(K+1)^{3/2}}{K}t^{-4}=f_{\rm SD}(K+1)^{3/2}\frac{dn_b}{dt}\,,
\label{loop_SD_rate}
\ee  
where $f_{\rm SD}$ is the fraction of newly born loops which 
die through the SD process. 

The rate of release of heavy right-handed neutrinos (we shall hereafter 
denote them by $N$; see section IV below) due to SD process can then be 
written as 
\be
\left(\frac{dn_N}{dt}\right)_{\rm SD}=N_N \frac{dn_{\rm SD}}{dt}
=N_N f_{\rm SD}\frac{1}{x^{2}}\left(\Gamma G \mu\right)^{-1} 
\frac{(K+1)^{3/2}}{K}t^{-4}\,. 
\label{N_SD_rate}
\ee 

\subsubsection{Quick death} 
Some fraction of the loops may be born in configurations with 
waves of high harmonic number. Such string loops have been 
shown~\cite{siemens&kibble.95} to have a high probability of 
self-intersecting. Ref.~\cite{siemens&kibble.95} gives the 
self-intersecting probability of a loop as 
\be 
P_{SI}=1-e^{-\alpha-\beta N}\,,
\label{self_int_prob}
\ee
where $\alpha=0.4$, $\beta=0.2$, and N is the harmonics number. 

A self-intersecting loop would break up into two or more 
smaller loops. The process of self-intersection leaves behind 
``kinks'' on the loops, which themselves represent high harmonic 
configurations. So, the daughter loops would also further 
split into smaller loops. If a loop does self-intersect, it 
must do so within its one oscillation period, since the 
motion of a loop is periodic. Under this circumstance, since 
smaller loops have smaller oscillation periods, it 
can be seen that a single initially large loop of length $L$ 
can break up into a debris of tiny loops of size $\eta^{-1}$ 
(at which point they turn into the constituent massive particles) 
on a time-scale $\sim L$. Equation (\ref{birth_length}) then 
implies that a loop born at the time $t$ in a high harmonic 
configuration decays, due to repeated self-intersection, into 
massive particles on a time scale $\tau_{QD}\sim K\Gamma G \mu t 
\ll H^{-1}(t)$. It is thus a ``quick death'' (QD) process --- 
the loops die essentially instantaneously (compared to 
cosmological time scale) as soon as they are formed. 
Equation (\ref{birth_rate}), therefore, directly gives the 
rate at which  loops die through this quick death process: 
\be
\frac{dn_{\rm QD}}{dt}=f_{\rm QD}\frac{dn_b}{dt}\,,
\label{loop_QD_rate}
\ee  
where $f_{\rm QD}$ is the fraction of newly born loops that 
undergo QD. 

Note that, since these loops at each stage self-intersect 
and break up into smaller loops before completing one oscillation, 
they would lose only a negligible amount of energy in 
gravitational radiation. Thus, almost the entire original 
energy of these loops would eventually come out in the 
form of massive particles. 

Assuming again, as we did in the SD case, that each segment of length 
$\sim w \sim \mu^{-1/2}$ of the loop yields a number $N_N\sim O(1)$ of 
heavy right-handed Majorana neutrinos, we can write, using equations 
(\ref{loop_QD_rate}), (\ref{birth_rate}) and (\ref{birth_length}), the 
rate of release of the $N$'s due to QD process as 
\be
\left(\frac{dn_N}{dt}\right)_{\rm QD}=N_N\, f_{\rm QD}\frac{1}{x^2}
\mu^{1/2} t^{-3}\,.
\label{N_QD_rate}
\ee

It is interesting to note here that if all loops were to die through 
this QD process, i.e., if we take $f_{\rm QD}=1$ in equations 
(\ref{loop_QD_rate}) and (\ref{N_QD_rate}), then the situation is in 
effect exactly equivalent to the microscopic loop formation scenario 
described by equation (\ref{birth_rate_AH}), although the primary loops 
themselves are formed with macroscopic size given by equation 
(\ref{birth_length}). 

While the important issue of whether or not massive particle radiation 
plays a dominant role in cosmic string evolution remains to be settled, 
the standard model may, of course, still allow a small but finite 
fraction, $f_{\rm QD}\ll 1$, of quickly dying loops. There already exist, 
however, rather stringent astrophysical 
constraints~\cite{pijush&rana,pijushreport} on 
$f_{\rm QD}$ from the observed flux of ultrahigh cosmic rays 
(UHECR) above $10^{11}\gev$~\cite{uhecr_obs} and the cosmic 
diffuse gamma ray background in the energy region 10 MeV -- 100 
GeV measured by the EGRET experiment~\cite{sreekumar}. This comes 
about in the following way: 

The massive $X$ particles released from the string loops would  
decay to SM quarks and leptons. The hadronization of the 
quarks gives rise to nucleons and pions with energy up to 
$\sim M_X$, the mass of the relevant $X$ particle. The neutral 
pions decay to photons. These 
extremely energetic nucleons and photons, after propagating 
through the cosmic radiation background, can survive as 
ultrahigh energy particles. The observed flux of UHECR, 
therefore, puts constraints on the rate of release of the 
massive $X$ particles, thereby constraining $f_{\rm QD}$. 
The most stringent constraint on $f_{\rm QD}$, however, comes 
from the fact that the electromagnetic component (consisting of 
photons and electrons/positrons) of the total energy injected 
in the Universe from the decay of the $X$ particles initiates 
an electromagnetic cascade process due to interaction of the 
high energy electrons/positrons and photons with the 
photons of the various cosmic background radiation fields (such as 
the radio, the microwave and the infrared/optical backgrounds); 
see, e.g., Ref.~\cite{pijushreport} for a review. As a result, 
a significant part of the total injected energy cascades down 
to lower energies. The measured flux of the cosmic gamma ray 
background in the 10 MeV -- 100 GeV energy region~\cite{sreekumar} 
then puts the constraint~\cite{pijushreport}
\be
f_{\rm QD}\eta_{16}^{2}\leq 9.6\times 10^{-6}\,, 
\label{f_QD_constraint}
\ee
where $\eta_{16}\equiv(\eta/10^{16}\gev)$. For GUT 
scale cosmic strings with $\eta_{16}=1$, for example, the above 
constraint implies that $f_{\rm QD}\leq 10^{-5}$, so that most loops 
should be in non-selfintersecting configurations, consistent with the 
standard scenario of cosmic string evolution.  
Note, however, that $f_{\rm QD}$ is not constrained by the above 
considerations for cosmic strings formed at a scale $\eta\lsim 
3.1\times 10^{13}\gev$. 

In this context, it is interesting to note that there is no 
equivalent constraint on the corresponding parameter 
$f_{\rm SD}$ for the slow death case from gamma ray background 
consideration. The reason is that, unlike in the QD case where 
the entire initial energy of a large loop goes into $X$ 
particles, only $\sim$ one $X$ particle is released from a 
initially large loop in the SD case. This in turn makes the 
time dependence of the rate of release of massive particles 
$\propto t^{-4}$ in the SD case (see equation (\ref{N_SD_rate})), while it 
is $\propto t^{-3}$ in the QD case (see equation (\ref{N_QD_rate})). Thus, 
while the SD process dominates at sufficiently early times, the QD process 
can dominate at relatively late times and can potentially contribute to 
the non-thermal gamma ray background. 

%%%%%%%%%%%%%%%%%%%%%%%
\section{Calculation of baryon asymmetry}
\subsection{Decay of heavy right-handed Majorana neutrinos and 
$L$-asymmetry}
The heavy right-handed Majorana neutrino decays to a SM lepton ($\ell$) 
and higgs ($\phi$) through the Yukawa coupling  
\begin{equation}
\mathcal{L}_{Y}=f_{ij}\bar{\ell}_i\phi\nu_{Rj}+h.c.\,,
\label{yukawa_term}
\end{equation}
where $f_{ij}$ is the Yukawa coupling matrix, and $i,j=1,2,3$ for three  
flavors.  

We shall work in a basis in which the right-handed Majorana neutrino mass 
matrix $M$ is diagonal, $M={\rm diag}(M_1,M_2,M_3)$. In this
basis the right-handed Majorana neutrino is given by 
$N_{j}=\nu_{Rj}\pm \nu^{c}_{Rj}$, which satisfies 
$N_{j}^{c}=\pm N_{j}$. The standard see-saw mechanism then gives the 
corresponding light neutrino mass eigenstates 
$\nu_1$, $\nu_2$, $\nu_3$ with masses $m_1$, $m_2$, $m_3$, respectively; 
these are mixtures of flavor eigenstates 
$\nu_e$, $\nu_\mu$, $\nu_\tau$, and are also Majorana neutrinos, i.e.,  
$\nu_i=\nu^c_i$.   

The decays of the heavy right-handed Majorana neutrinos can create a 
non-zero $L-asymmetry$ (which is ultimately converted  
to $B$-asymmetry) only if their decay violates CP. 
The CP-asymmetry parameter in the decay of $N_j$ is defined as  
\be 
\epsilon_{j}\equiv\frac{\Gamma(N_j\rightarrow \ell\phi)-
\Gamma(N_j\rightarrow \ell^c\phi^c)}{\Gamma(N_{j}
\rightarrow \ell\phi)+\Gamma(N_{j}\rightarrow \ell^{c}
\phi^{c})}\,.
\label{epsilon_def}
\ee

Assuming a mass hierarchy in the heavy neutrino sector, 
$M_{1} < M_{2} < M_{3}$, it is reasonable to expect that the  
final lepton asymmetry is produced mainly by the decay of the 
lightest right handed neutrino $N_{1}$. Any asymmetry produced by the 
the decay of $N_{2}$ and $N_{3}$ will be washed out by the lepton number 
violating interactions mediated by the $N_{1}$. As the Universe 
expands, the temperature of the thermal plasma falls. Below a 
temperature $T_F\sim M_1$, all $L$-violating scatterings mediated by 
$N_{1}$ freeze out, thus providing the out-of-equilibrium 
situation~\cite{sakharov.67} necessary for the survival of any net 
$L$-asymmetry generated by the decay of the $N_1$'s. 
The final $L$-asymmetry is, therefore, given essentially by the 
CP asymmetry parameter $\epsilon_1$.  

An accurate calculation of the net $L$-asymmetry can only be done by 
numerically solving the full Boltzmann equation that includes all lepton 
number violating interactions involving all the $N_j$'s present at any 
time, including the $N_j$'s of non-thermal origin such as the ones 
produced from the decaying cosmic string loops, as well as those of 
thermal origin. This is beyond the scope of 
the present paper; here we shall simply assume that below the temperature 
$T_F=M_1$, all interactions except the decay of the $N_1$ are unimportant, 
so that each $N_1$ released from cosmic strings additively produces a net 
$L$-asymmetry $\epsilon_1$ when it decays. 

To fix the value of $\epsilon_1$, we note that there is an upper 
bound~\cite{davidson_ibarra} on $\epsilon_1$, which is related to the 
properties of the light neutrino masses. In a standard hierarchical 
neutrino mass scenario with $m_3\gg m_2>m_1$, this upper limit is given 
by~\cite{davidson_ibarra,buch_bari_plum} 

\be
|\epsilon_{1}|\leq \frac{3}{16\pi}\frac{M_1 m_3}{v^2}\,, 
\label{epsilon_1_max}
\ee 
where $v\simeq 174\gev$ is the electroweak symmetry breaking scale. 
Furthermore, the above upper limit is in fact 
{\it saturated}~\cite{buch_bari_plum} in most of the reasonable neutrino 
mass models, which we shall assume to be the case. 

The atmospheric neutrino data~\cite{atmos_data} indicate 
$\nu_\mu\leftrightarrow \nu_\tau$ oscillations with nearly maximal mixing 
($\theta_{\rm atm}\simeq 45^\circ$) and a mass-squared-difference  
$\Delta m^2_{\rm atm}\equiv |m_3^2-m_2^2|\approx 2.6\times10^{-3}\ev^2$.  
The solar neutrino~\cite{solar_data} and KamLAND~\cite{kamland_data}
data, on the 
other hand, can be explained by $\nu_e\leftrightarrow \nu_\mu$ 
oscillations with large mixing angle (LMA) ($\theta_{\rm sol}\simeq 
32^\circ$) and 
$\Delta m^2_{\rm sol}\equiv |m_2^2-m_1^2|\approx 7.13\times10^{-5}\ev^2$. 
Assuming, again, the standard light neutrino mass 
hierarchy, the above numbers give $m_3\simeq \left(\Delta m^2_{\rm 
atm}\right)^{1/2}\simeq 0.05\ev$. In our calculations below, we shall use 
\be
\epsilon_1\simeq 9.86\times 10^{-4}\left(\frac{M_1}{10^{13}\gev}\right)
\left(\frac{\left(\Delta m^2_{\rm atm}\right)^{1/2}}{0.05\ev}\right)\,.
\label{epsilon_1_value}
\ee 

The $L$-asymmetry is partially converted to a $B$-asymmetry by the rapid  
nonperturbative sphaleron transitions which violate $B+L$ but preserve 
$B-L$. Assuming that sphaleron transitions are ineffective at temperatures 
below the electroweak transition temperature ($T_{\rm EW}$), the 
$B$-asymmetry is related to $L$-asymmetry by the 
relation~\cite{harvey_turner} 
\be
B=p\, (B-L)=\frac{p}{p-1}L\simeq -0.55 L\,,
\label{B_L_relation}
\ee
where we have taken $p=28/79$ appropriate for the particle 
content in SM~\cite{harvey_turner}. If sphaleron transitions continue to 
be effective below $T_{\rm EW}$, then the above relation between $B$ and 
$L$ is slightly modified; we can however ignore this at the level of 
accuracy aimed at in the present paper.  

The net baryon asymmetry of the Universe is defined as 
\be
Y_B=\frac{n_B-n_{\bar{B}}}{s}\,,
\label{bau_def}
\ee
where 
\be 
s\simeq (2\pi^2/45)g_* T^3\simeq 43.86 \left(\frac{g_*}{100}\right) T^3
\label{entropy_density}
\ee  
is the entropy density, $g_*$ being  
the number of relativistic degrees of freedom contributing to the entropy 
at the temperature $T$. At temperatures in the early Universe relevant 
for the process of baryon asymmetry generation, $g_*\simeq 100$ in SM.      

Observationally, the BAU is often expressed in terms of the  
baryon-to-photon ratio $\eta\equiv (n_B-n_{\bar{B}})/n_\gamma$, whose 
present-day-value $\eta_0$ is related to that of $Y_B$ through the 
relation
\be
\eta_0\simeq 7.0 Y_{B,0}\,.
\label{eta_0_def}
\ee
The observed value of $\eta_0$ inferred from the Wilkinson Microwave 
Anisotropy Probe (WMAP) data is~\cite{wmap} 
\be
\eta_0^{\rm WMAP}=\left(6.1^{+0.3}_{-0.2}\right)
\times 10^{-10}\,. 
\label{eta_0_wmap}
\ee 

We now proceed to estimate the contribution to the BAU from the two 
cosmic string loop processes discussed in the previous section. 

\subsection{Slow death case}
The contribution of the SD process to $\eta_0$ can be written as    
\be
\eta_0^{\rm SD}\simeq 7.0 \times 0.55\, \epsilon_1 
\int_{t_{F}}^{t_{0}}\frac{1}{s} 
\left(\frac{dn_N}{dt}\right)_{\rm SD} dt\,,
\label{eta_0_SD_def}
\ee 
where $t_F$ is the cosmic time corresponding to the temperature $T_F\simeq 
M_1$ and $t_0$ is the present age of the Universe. Using 
equations (\ref{N_SD_rate}), (\ref{entropy_density}) and the standard 
time-temperature relation in the early Universe,  
\be
t\simeq 0.3 g_*^{-1/2}\frac{M_{\rm Pl}}{T^2}\,,
\label{t_T_rel}
\ee 
where $M_{\rm Pl}\simeq 1.22\times 10^{19}\gev$ is the Planck mass, we see 
that the dominant contribution to the integral in equation 
(\ref{eta_0_SD_def}) comes from 
the time $t_F\ll t_0$, i.e., from the epoch of temperature $T_F\simeq 
M_1$, giving 
\bea
\eta_0^{\rm SD} & \simeq 2.0\times 10^{-7} N_N    
\left(\frac{M_{1}}{10^{13}\gev}\right)^{4} 
\left(\frac{\eta_{B-L}}{10^{13}\gev}\right)^{-2}\nonumber \\
& =2.0\times 10^{-7} N_N h_1^4 
\left(\frac{\eta_{B-L}}{10^{13}\gev}\right)^{2}\,,
\label{eta_0_SD_value}
\eea 
where we have defined the Yukawa coupling $h_1\equiv M_1/\eta_{B-L}$, 
used equation (\ref{epsilon_1_value}) for $\epsilon_1$ 
with $\left(\Delta m^2_{\rm atm}\right)^{1/2} = 0.05\ev$, and also 
taken $x=0.5$, $\Gamma=100$, $K=1$ and $f_{\rm SD}=1$ in 
equation (\ref{N_SD_rate}). 

The Yukawa couplings are generally thought to be less than unity. 
With $h_1\leq 1$, we see from (\ref{eta_0_SD_value}) and 
(\ref{eta_0_wmap}) that the cosmic 
string loop slow death process can produce the observed BAU 
only for $B-L$ phase transition scale 
\be 
\eta_{B-L}^{\rm SD}\gsim 5.5\times 10^{11}N_N^{-1/2}\gev\,. 
\label{eta_scale_lower_limit_SD}
\ee  
Assuming $N_N\lsim 10$, say, we see that cosmic string loop SD process can 
contribute to BAU if $\eta_{B-L}\gsim 1.7\times10^{11}\gev$; lower values 
of $\eta_{B-L}$ are relevant only if we allow $h_1>1$. 

At the same time, for a given $\eta_{B-L}$ satisfying 
(\ref{eta_scale_lower_limit_SD}), in order that 
the contribution (\ref{eta_0_SD_value}) not exceed the 
highest allowed observed value of $\eta_0$ given by equation 
(\ref{eta_0_wmap}), the Yukawa coupling $h_1$ must satisfy the constraint   
\be
h_1^{\rm SD}\lsim 0.24 
\left(\frac{\eta_{B-L}}{10^{13}\gev}\right)^{-1/2}N_N^{-1/4}\,,
\label{h_1_SD_constraint}
\ee 
which, in terms of the lightest heavy right handed Majorana 
neutrino mass $M_1$, reads 
\be
M_1^{\rm SD}\lsim 2.4 \times 10^{12} N_N^{-1/4}  
\left(\frac{\eta_{B-L}}{10^{13}\gev}\right)^{1/2}\gev\,.
\label{M1_constraint_SD}
\ee
Note the rather weak dependence of the above constraints on $N_N$. 
Also, the 4th power dependence on $M_1$ of equation 
(\ref{eta_0_SD_value}) and the rather narrow range of the observed value 
of $\eta_0$ given by equation (\ref{eta_0_wmap}) together imply that, in 
order for the SD process to explain the observed BAU, $M_1$ (and 
equivalently $h_1$) cannot be much 
smaller than their respective values saturating the above constraints. 

\subsection{Quick death case}
Replacing $\left(\frac{dn_N}{dt}\right)_{\rm SD}$ in equation 
(\ref{eta_0_SD_def}) by $\left(\frac{dn_N}{dt}\right)_{\rm QD}$ given by 
equation (\ref{N_QD_rate}), and following the same steps as in the SD case 
above, we get the contribution of the QD process to $\eta_0$ as 
\bea
\eta_0^{\rm QD} & \simeq 5.17 \times 10^{-13}N_N f_{\rm QD} 
\left(\frac{M_1}{10^{13}\gev}\right)^2
\left(\frac{\eta_{B-L}}{10^{13}GeV}\right)\nonumber \\
& = 5.17\times 10^{-13}N_N f_{\rm QD} h_1^2
\left(\frac{\eta_{B-L}}{10^{13}GeV}\right)^3\,.
\label{eta_0_QD_value_1}
\eea
From (\ref{eta_0_QD_value_1}) and (\ref{eta_0_wmap}) we see that, 
considering the most optimistic situation with $f_{\rm QD}=1$, the QD 
process is relevant for BAU only for 
\be 
\eta_{B-L}^{\rm QD}\gsim 1.1\times 10^{14}N_N^{-1/3}\gev\,; 
\label{eta_scale_lower_limit_QD_1}
\ee  
lower values of $\eta_{B-L}$ are relevant only if we allow $h_1>1$. On the 
other hand, the constraint 
(\ref{f_QD_constraint}) allows $f_{\rm 
QD}=1$ {\it only if} $\eta_{B-L}\leq 3.1\times10^{13}\gev$. This can be 
reconciled with the above constraint (\ref{eta_scale_lower_limit_QD_1}) 
only for $N_N > 45$ or so. Such a large value of $N_N$ seems unlikely.  

In general, using the constraint (\ref{f_QD_constraint}) on 
$f_{\rm QD}$ in (\ref{eta_0_QD_value_1}) we get 
\bea
\eta_0^{\rm QD} & \lsim 5.0 \times 10^{-12}N_N  
\left(\frac{M_1}{10^{13}\gev}\right)^2
\left(\frac{\eta_{B-L}}{10^{13}GeV}\right)^{-1}\nonumber \\
& = 5.0\times 10^{-12}N_N h_1^2
\left(\frac{\eta_{B-L}}{10^{13}GeV}\right)\,.
\label{eta_0_QD_value_2}
\eea
Comparing again with the observed value of $\eta_0$, we now 
see that, for $h_1\leq 1$, the QD process can be relevant for BAU only 
for 
\be 
\eta_{B-L}^{\rm QD}\gsim 1.2\times 10^{15}N_N^{-1}\gev\,. 
\label{eta_scale_lower_limit_QD_2}
\ee  

For values of $\eta_{B-L}$ satisfying the above constraint 
(\ref{eta_scale_lower_limit_QD_2}), the QD process can produce the 
observed value of BAU for 
\be
h_1^{\rm QD}\lsim 0.36
\left(\frac{\eta_{B-L}}{10^{16}\gev}\right)^{-1/2}N_N^{-1/2}\,,
\label{h_1_QD_constraint}
\ee
which in terms of $M_1$ now reads
\be
M_1^{\rm QD}\lsim 3.6 \times 10^{15} N_N^{-1/2}
\left(\frac{\eta_{B-L}}{10^{16}\gev}\right)^{1/2}\gev\,.
\label{M1_constraint_QD}
\ee

From the above discussions we see that, as far as their contributions to 
the BAU is concerned, the QD process becomes important only at relatively 
higher values of the symmetry breaking scale $\eta_{B-L}$ compared to the 
SD process. 

%%%%%%%%%%%%%%%%%%%%%%%%%%%%%%%
\section{Summary and conclusions}
A wide class of unified theories with an embedded $U(1)_{B-L}$ gauge 
symmetry allows formation of ``$B-L$'' cosmic strings at the $U(1)_{B-L}$ 
symmetry-breaking phase transition at a symmetry-breaking scale 
$\eta_{B-L}$ well below the GUT scale ($\sim 10^{16}\gev$). Such ``light'' 
cosmic strings are not currently excluded by CMB anisotropy data.
The $B-L$ cosmic strings can carry zero modes of heavy 
right handed Majorana neutrinos ($N$), and the latter can be released 
from closed loops of these cosmic strings when the loops eventually 
disappear. The decay of the $N$'s can give rise to a $L$-asymmetry which 
is partially converted to $B$-asymmetry via nonperturbative sphaleron 
transitions. 

In this paper we have studied the contribution to the baryon asymmetry of 
the Universe due to decay of heavy right handed Majorana neutrinos 
released from closed loops of $B-L$ cosmic strings in the light of current 
ideas on light neutrino masses and mixings implied by atmospheric 
and solar neutrino measurements. We have estimated the 
contribution to BAU from cosmic string loops which disappear through the  
process of (a) slow shrinkage due to energy loss through gravitational 
radiation --- which we call slow death (SD), and (b) repeated 
self-intersections --- which we call quick death (QD). We find that 
for reasonable values of the relevant parameters, the SD process 
dominates over the QD process as far as their contribution to BAU is 
concerned. We find that $B-L$ cosmic string loop SD process can 
contribute significantly to, and can in principle produce, the observed 
BAU for $U(1)_{B-L}$ symmetry breaking scale $\eta_{B-L}\gsim 1.7\times 
10^{11}\gev$. The QD process, on the other hand, becomes relevant for BAU 
only for relatively higher values of $\eta_{B-L}\gsim 10^{14}\gev$. 

We have also found that, in order for the contribution of 
decaying $B-L$ cosmic string loop SD process not to exceed the 
observed baryon-to-photon ratio inferred from the recent WMAP 
results, the lightest heavy right-handed Majorana neutrino 
mass $M_1$ must satisfy the constraint $M_1 \leq 2.4 \times 
10^{12}\left(\eta_{B-L}/10^{13}\gev\right)^{1/2}\gev$. 
This result may have interesting implications for the 
associated Yukawa couplings in the heavy neutrino sector and consequently 
for the light neutrino masses generated through see-saw mechanism. 

We conclude that processes involving closed loops of cosmic strings formed 
at a $U(1)_{B-L}$ symmetry breaking phase transition may make 
significant contribution to the observed BAU and should be included in 
considerations of baryon generation processes in general. A full Boltzmann 
equation calculation of the baryogenesis process including the heavy right 
handed neutrinos of cosmic string origin as well of thermal origin is in 
progress and will be reported elsewhere.  

%%%%%%%%%%%%%%%%%%%%%%%%%%%%%%%%%%%%%

\end{document}